**Chapter 12**

# HCI in E-Government and E-Democracy

Tianmu Zhu [0009-0002-6569-5969]

Zhejiang Sci-Tech University

Hangzhou, China

1303145256@qq.com

Wei Xu [0000-0001-8913-2672]

Zhejiang University

Hangzhou University

weixu6@yahoo.com

**Abstract:** This chapter introduces the application of HCI design processes and design principles in the context of e-government and e-democracy. We elaborate on HCI design processes and six HCI design principles in the context of e-government and e-democracy, including citizen-centered design, usability, accessibility, access to information, transaction efficiency, and security and privacy. Then, we present three cases to demonstrate the value of applying HCI processes and design principles in developing and deploying e-government and e-democracy. Finally, we highlight emerging challenges for e-government and e-democracy as well as future trends. In conclusion, HCI can help the success of e-government and e-democracy and their future growth.

## 1   Introduction

E-government and e-democracy have achieved vigorous development on a worldwide scale. Combined with advanced Internet and information technology and popularized electronic devices' support, citizens could have a one-stop operation on e-government websites instead of physically visiting government departments premises. E-government refers to delivering information and services from a national or local government to stakeholders (e.g., citizens, businesses, and other governments) through the Internet or other digital platforms. E-democracy refers to using digital communication technologies, particularly the Internet, to enhance and facilitate democratic processes and engagement, especially online voting (e-voting). E-democracy is an integral part of e-government. E-government emphasizes the delivery of information, whereas e-democracy emphasizes citizen engagement.

The widely developed e-commerce paradigm has benefited the implementation of e-government and e-democracy applications. The e-commerce paradigm has brought practical lessons for e-government and e-democracy, but it is not appropriate to apply it directly due to different user needs. This chapter elaborates on six HCI design principles, including citizen-centered design, usability, accessibility, access to information, transaction efficiency, and security and privacy. HCI provides powerful support for the exquisite and thriving development of e-government and e-democracy. The application of HCI is demonstrated through two examples: the e-government website (called "Yi Wang Tong Ban" in Chinese) implemented by the Shanghai government (China) and an e-voting system, namely STAR-vote.

In addition, challenges always emerge in the development process of innovation, and challenges bring opportunities simultaneously, as in e-government and e-democracy, such as multilingual and multicultural inclusivity. Also, future trends are emerging, such as personalization and customization, the application of new technologies with big data, blockchain, and artificial intelligence (AI).

The content of this chapter is organized as follows. We first introduce the three concepts e-government, e-democracy, and the e-commerce paradigm; then we elaborate on HCI design processes and six HCI design principles in the context of e-government and e-democracy. To illustrate the application of HCI, we present three case studies by assessing how HCI methods and design principles were applied to improve the design of an e-government website and an e-democracy website. Finally, we highlight the challenges faced in applying HCI in e-government and e-democracy and emerging future trends.

## 2 E-government and E-democracy

### 2.1 E-government

E-government (electronic government) originates from developing electronic information technology and transforming public management concepts. It is an essential component of modern national governance systems and governance capabilities. E-government refers to delivering information and services from a national or local government to stakeholders (e.g., citizens, businesses, and other governments) through the Internet or other digital platforms. Although the definitions of e-government differ slightly across organizations, the main goal shared across organizations is using information and communication technologies (ICTs) to improve governance (Palvia & Sharma, 2007).

The e-government development index (EGDI) was developed by the United Nations Department of Economic and Social Affairs (UNDESA) of the United Nations to assess the status of countries and regions in terms of e-government maturity worldwide (UNPAN, 2022). Figure 1 shows the map of EGDI value groups from 193 United Nations Member States. Dark blue represents the countries with very high EGDI values (0.75-1.00), blue represents high (0.50-0.75), sky blue represents middle (0.25-0.50), and light blue represents low (0.00-0.25). In 2022, 60 countries have very high EGDI values, increased by three compared to 2020. The quantity of the other three groups in order is 73, 53, and 7.

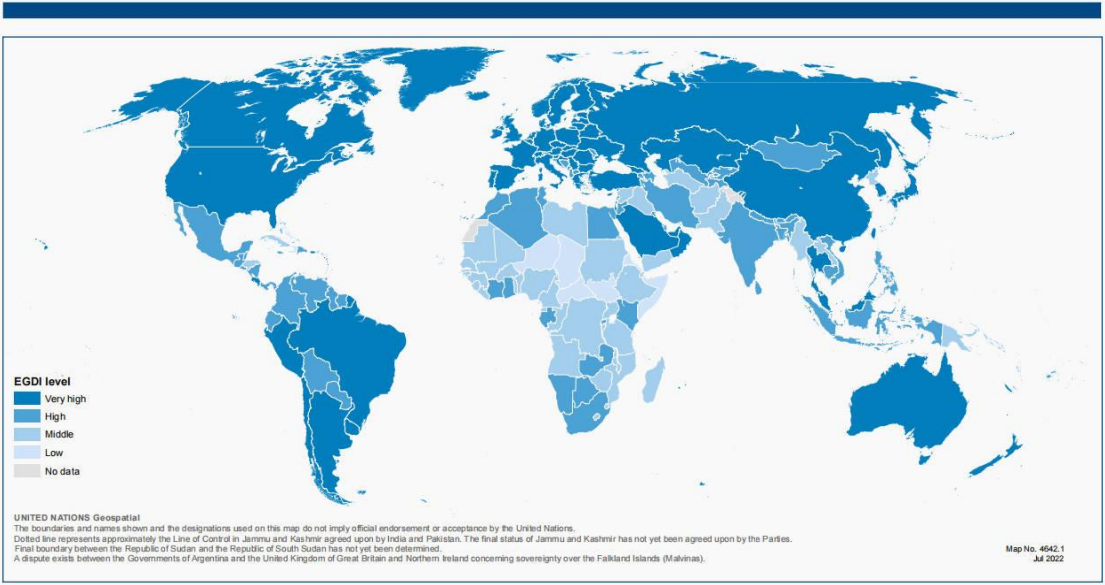

Fig. 1. Geographical distribution of the four EGDI groups (United Nations E-Government Survey 2022)

The COVID-19 pandemic has amplified the significant role of e-government in all aspects of digital life. In response to COVID-19, e-governments offered four measures (see Figure 2). As shown in Figure 2, almost 90% of countries in Europe implemented these four measures during the pandemic. In contrast, less than 50% of the countries in Africa and Oceania implemented the four measures. Americas and Asia performed second to Europe. This survey illustrates that e-government has played a key role, especially in unexpected events. It is necessary to reflect and review the deficiencies to prepare for subsequent unexpected events.

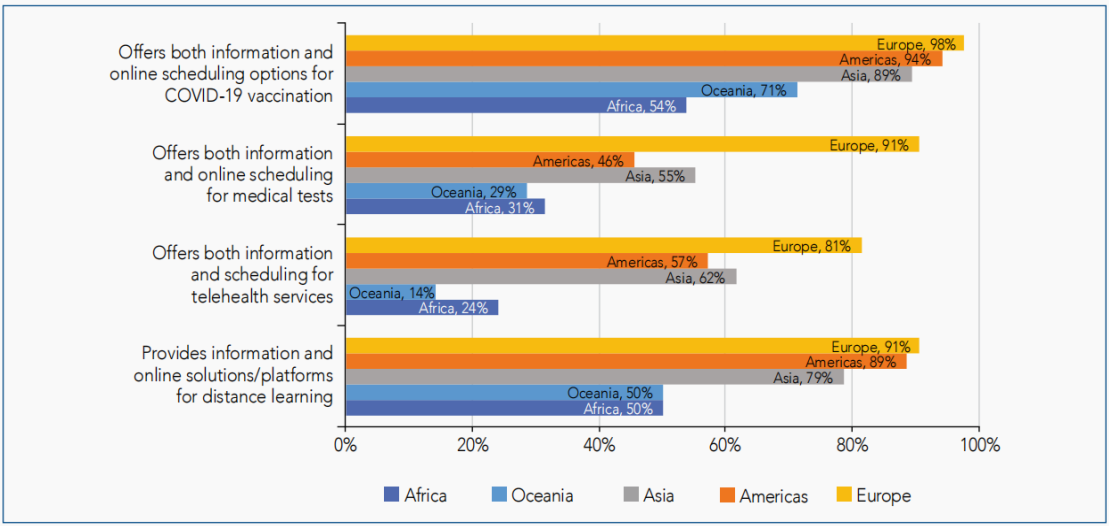

Fig. 2. The implementation of COVID-19 response measures by region (United Nations E-Government Survey 2022)

UNDESA also proposed LOSI (Local Online Service Index) to reflect the state of local e-government development at city level. It is the first time for LOSI to incorporate an assessment of e-government in the most populous city in each of the 193 member countries in the 2022 edition of the United Nations E-Government Survey. Table 1 shows the results with high LOSI scores (0.75-1.00). Geographically, 20 of the 38 cities are located in Europe, 10 in Asia, 6 in the Americas, and 2 in Oceania. The scores of some cities are relatively close or tied, such as Berlin and Madrid tied for first place, and Dubai, Moscow,

New York, and Paris tied for fifth place. Shanghai ranks 10th of the 38 cities, and Shanghai e-government Yi Wang Tong Ban will be discussed as a case in section 4.1.

Table 1: Cities with high LOSI scores (United Nations E-Government Survey 2022)

| City | Country | Rank | LOSI score | City | Country | Rank | LOSI score |
|---|---|---|---|---|---|---|---|
| Berlin | Germany | 1 | 0.9767 | Reykjavik | Iceland | 19 | 0.8372 |
| Madrid | Spain | 1 | 0.9767 | Helsinki | Finland | 21 | 0.8256 |
| Tallinn | Estonia | 3 | 0.9535 | Kiev | Ukraine | 21 | 0.8256 |
| Copenhagen | Denmark | 4 | 0.9419 | Riga | Latvia | 21 | 0.8256 |
| Dubai | United Arab Emirates | 5 | 0.9186 | Stockholm | Sweden | 21 | 0.8256 |
| Moscow | Russian Federation | 5 | 0.9186 | Manama | Bahrain | 25 | 0.8140 |
| New York | United States of America | 5 | 0.9186 | Almaty | Kazakhstan | 26 | 0.8023 |
| Paris | France | 5 | 0.9186 | Luxembourg City | Luxembourg | 26 | 0.8023 |
| Singapore | Singapore | 9 | 0.9070 | Vilnius | Lithuania | 28 | 0.8023 |
| Shanghai | China | 10 | 0.8837 | Montevideo | Uruguay | 29 | 0.7907 |
| Bogota | Colombia | 11 | 0.8721 | Seou | Republic of Korea | 30 | 0.7674 |
| Buenos Aires | Argentina | 11 | 0.8721 | Tel Aviv | Israel | 30 | 0.7674 |
| Istanbul | Turkiye | 11 | 0.8721 | Toronto | Canada | 30 | 0.7674 |
| Tokyo | Japan | 14 | 0.8605 | Warsaw | Poland | 30 | 0.7674 |
| Zurich | Switzerland | 14 | 0.8605 | Brussels | Belgium | 34 | 0.7558 |

| Rome | Italy | 16 | 0.8488 | Oslo | Norway | 34 | 0.7558 |
| Sao Paulo | Brazi | 16 | 0.8488 | Riyadh | Saudi Arabia | 34 | 0.7558 |
| Vienna | Austria | 16 | 0.8488 | Sydney | Australia | 34 | 0.7558 |
| Auckland | New Zealand | 19 | 0.8372 | Zagreb | Croatia | 34 | 0.7558 |

## 2.2 E-democracy

E-democracy refers to using ICT to support democratic decision-making, which enhances democratic institutions and democratic processes (Ronchi, 2019). E-engagement means engaging the public in the policy process via electronic networks (Palvia & Sharma, 2007). E-democracy is not a substitute for traditional forms of democracy but rather a means to make democratic institutions more efficient and productive. Abu-Shanab (2015) proposed an e-government framework (see Figure 3) including four major dimensions of e-government: service provision, government performance, democracy (political side of e-government), and the social contribution of technology. E-democracy activities included e-elections, e-voting, e-petitioning, e-consultation, and e-participation (Abu-Shanab, 2015). E-democracy is an integral part of e-government and a sub-discipline of e-government research (Medaglia, 2007).

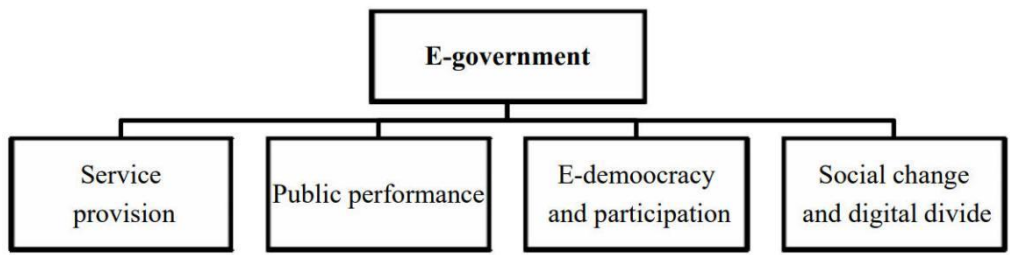

Fig. 3. E-government framework (Abu-Shanab, 2015)

Unlike the above model, Roztocki et al. (2023) proposed a comprehensive e-democracy framework encompassing e-participation, e-government, e-justice, and e-parliament, with particular emphasis on e-participation (see Figure 4). E-participation permeates all of e-democracy, so it is at the top of the framework. In this framework, e-democracy extends beyond, enabling the broader population to participate in democratic processes (e-inclusion) and engage in policy-making (e-participation). E-government remains responsible for transmitting information and providing services, e-justice reduces the cost and burden of the judicial system, and e-parliament is the utilization of ICT in the performance of legislative function (Roztocki et al., 2022).

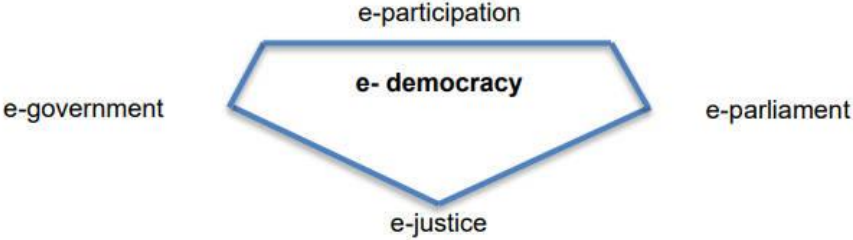

Fig. 4. E-democracy framework (Roztocki et al., 2023)

## 2.3 E-commerce paradigm

An e-commerce paradigm refers to buying and selling products, services, or other commodities through information and communication technology (Stahl, 2005a). Because of the three advantages of the e-commerce paradigm (i.e., efficiency, cost-saving, and customer-centeredness), the e-commerce paradigm developed successfully in the past decades. Efficiency is the primary reason why the e-commerce paradigm has been used in e-government and e-democracy. E-government aims at efficient operations, whereas government bureaucracies are inefficient. E-democracy aims to make citizen participation more convenient and effective.

However, there are similarities and differences between e-commerce, e-government, and e-democracy. Scholl et al. (2009) summarized the similarities and differences in five aspects: information management, process management, stakeholder relations, customer/citizen focus, and digital divide. For information management, both e-commerce and e-government faced challenges in managing information volume, accuracy, and timeliness. Both emphasize ease of use and usability. For process management, e-commerce had more advanced process redesign compared to e-government. While transaction speed was crucial in e-commerce, it received less emphasis in e-government. For stakeholder relations, balancing stakeholder interests and managing expectations were similar in both sectors, but e-government requires more emphasis on public decision-making. For customer/citizen focus, e-commerce was driven by profits and customer services, while e-government focused on serving citizens and encouraging participation. For the digital divide, e-government was concerned about the digital divide, including issues like equal access, literacy, reach, language, content, and infrastructure; e-commerce was less concerned about the digital divide, primarily focusing on reaching potential customers and providing secure transactional spaces.

Moreover, there are differences between customers, citizens, companies, and governments. Citizens are not only customers but also the e-government and e-democracy shareholders. Furthermore, market competition does not exist among citizens and governments (Stahl, 2005b). Generally, citizens cannot easily choose e-government websites, and it is hard for citizens to change the government they belong to unless they migrate. On the contrary, customers can always choose different products or companies. If the chosen one does not satisfy them, they can choose another. Citizens have the right to get information and services from e-government websites, and there is no need to compete with others. Thus, e-government and e-democracy have unique characteristics compared to e-Commernce.

## 2.4 Why do we need HCI in e-government and e-democracy?

The application of HCI in e-commerce started relatively early and is now fairly widespread. A number of HCI studies in e-commerce can be mentioned, for example, web usability, user interface (UI) design, navigation, and trust (Nah, 2002; Xu, 2005, 2017), trust issues for e-commerce system design (Egger, 2000), the UI design for cyber shopping Malls (Kim, 1997), and cross-disciplinary HCI research for customer relationship marketing (Petre et al., 2006). However, the implementation of HCI design principles and processes started relatively late in the application of e-government and e-democracy, and the HCI aspect was not considered at the beginning of the design (Olembo & Volkamer, 2013; Acemyan et al., 2022). Many studies have revealed that the usability of e-government and e-democracy can be optimized to serve citizens better (Karayumak et al., 2011; Lyzara et al., 2019; Paul & Das, 2020; Eilu et al., 2021).

As mentioned, two obvious shortcomings exist in e-government and e-democracy. Firstly, e-government and e-democracy do not emphasize the application of HCI design principles and processes in websites or systems. For instance, early prototyping of the user interface (UI) and usability evaluation could identify user experience (UX) issues before release. Secondly, there is a lack of widespread application of HCI design principles and processes to achieve fundamental UX objectives as done in e-commerce. Moreover, when implementing user-centered design, attention should also be paid to the differences in user roles between e-commerce and citizens. HCI could solve the two problems through UI prototyping and usability testing, applying HCI design principles and guidelines, as well as other methods. HCI focuses on user-centered design; in e-government and e-democracy, this entails citizen-centered design. HCI calls for careful consideration of the unique characteristics and needs of users from the perspective of citizens, addressing their needs and UX issues in the UI design of e-government and e-democracy websites/applications during the development process, validating their needs by testing the proposed websites/applications, and continued improvements of the websites/applications with user feedback over time.

The following section introduces six HCI design principles for e-government and e-democracy: (1) user-centered design/citizen-centered design, (2) usability, (3) accessibility, (4) access to information, (5) transaction efficiency, and (6) security and privacy. Applying the HCI principles and methods will make the e-government and e-democracy systems more efficient and effective. Also, HCI will enhance the overall performance of e-government and e-democracy beyond their websites and applications, further improving the efficiency and popularity of e-government and e-democracy and facilitating citizens to receive the best service.

## 3   HCI design principles and research in e-Government and e-Democracy

### 3.1   User-centered /citizen-centered design

HCI advocates user-centered design and emphasizes designing systems with users in mind (Xu, 2018). User-centered design means citizen-centered design in e-government and e-democracy, where citizens interact with digital platforms. A user-centered design approach ensures that these systems are intuitive, accessible, and meet the needs of diverse users. In ISO 9241-210: 2019 (en), human-centered design is defined as an "approach to systems design and development that aims to make interactive systems more usable by focusing on the use of the system and applying human factors/ergonomics and usability knowledge and techniques."

It was still a big challenge for successful applications for e-government; many projects failed because they did not meet the needs and skills of end users (Saqib & Abdus, 2018). Conventional e-government focuses on automation processes and efficient delivery of services; there is an increasing need for a user-centered design approach. The design of e-government should consider the real use scenarios of applying e-government, including factors such as the characteristics and cultural characteristics of citizens, and carry out targeted design centered around citizens rather than just designing a website or system to popularize e-government and electronic democracy (Prabhu & Raja, 2023).

For example, PortNL is an integrated service portal prototype for people who were temporarily relocated to live and work in the Netherlands. Figure 5 shows the PortNL website development process, which followed the ISO standard for user-centered design cycle processes (Kotamraju & van der Geest, 2012).

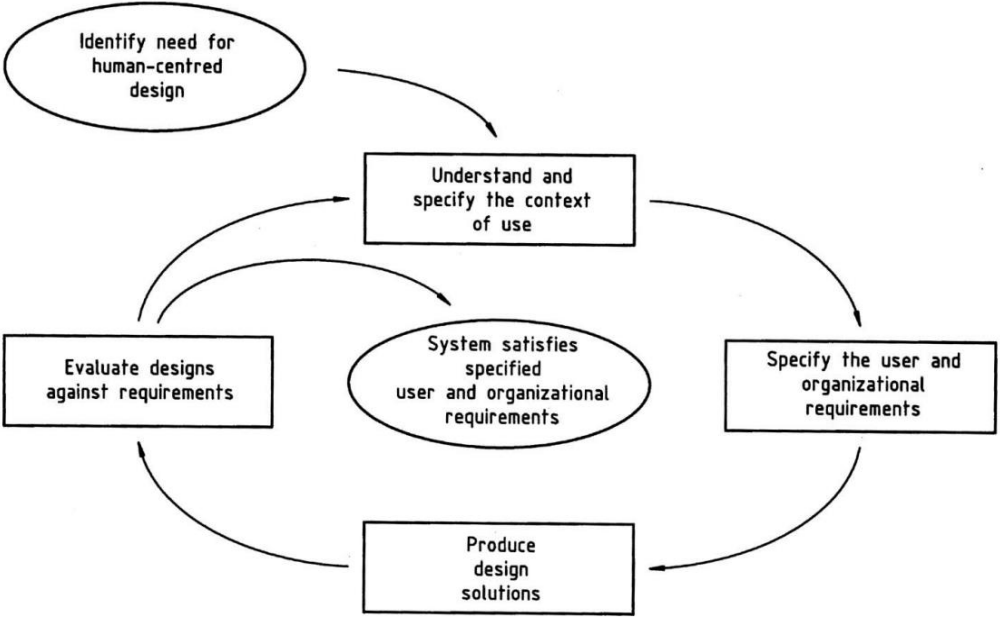

Fig. 5. User-centred design activities (Kotamraju & van der Geest, 2012)

Figure 5 shows several user-centered design activities conducted in the design process (Kotamraju & van der Geest, 2012). First, a pre-design study of 14 in-depth interviews was conducted with expatriates, investigating their official interactions with the Netherlands Government and paperwork before arrival. Second, users' expectations, preferences, perceptions, and other important factors were defined based on participants' feedback collected using a paper-based prototype of the Web portal. Third, critical reviews of the user interface design were conducted through users' self-reports, stakeholder meetings, and email discussions. In addition, researchers identified four forms of tensions between users and government:

(1) users and government hold contradictory visions of the tasks to be accomplished;

(2) governments must design for exceptions as well as for the mainstream;

(3) users and government differ in their commitments to legal rules and regulations;

(4) desires about the nature of users – government relationship are conflicting.

User-centered design is essential in addressing the tense relationship between citizens and government needs, iteratively solving problems by focusing on this conflicting relationship. Unlike service quality measurement, which often relies on post-service satisfaction assessments, user-centered design, emphasizing pre-design research and iterative design with users, offers stakeholders the necessary information for service design. For instance, Kotamraju & van der Geest (2012) gained insights from pre-design interviews, such as the challenges people faced when moving to the Netherlands without identification numbers, which satisfaction surveys might miss. For e-government services to be successful and effectively meet the needs of both users and government, it is crucial to enhance our understanding and application of user-centered design in the context of e-government. If successful, this effort will improve future government services and have broader applications in any large-scale, administratively complex, and public domain.

### 3.2 Usability

Usability is a fundamental and important principle in user-centered design. The International Organization for Standardization (ISO) defined usability as "the extent to which a system, product or service can be used by specified users to achieve specified goals with effectiveness, efficiency, and satisfaction in a specified context of use" (ISO 9241-11:2018) (Xu, 2003).

User-friendly interface design is very suitable for designing government websites because it takes both usability and UCD into profound consideration to facilitate citizens' use and adoption. However, user-friendly interface design and other HCI design principles or guidelines did not receive sufficient attention and application in the initial design of e-government websites. Many studies on the usability of e-government websites have revealed severe weaknesses; the interfaces and websites are not truly user-friendly (Garcia et al., 2005; Granizo et al., 2011; Karayumak et al., 2011; Lyzara et al., 2019; Paul & Das, 2020; Eilu et al., 2021).

Nielsen's 10 Usability Heuristics for UI Design (Nielsen, 1994) is a well-known guideline for evaluation in HCI and has been widely used. Considering the specificities of e-government websites, Garcia et al. (2005) proposed a g-Quality evaluation method as an extension of Nielsen's 10 Usability Heuristics. Studies found the g-Quality evaluation method could identify more usability problems than Nielsen's when evaluating 127 Brazilian e-gov sites (Garcia et al., 2005).

More specifically, the heuristics are grouped under five evaluation criteria in terms of the citizen-centered design approach (Garcia et al., 2005):

- Cognitive effort: minimizing the cognitive effort when citizens conduct a task, making sure the process is more intuitive and effective;
- Tolerance: citizens' motivation, patience, understanding, and performance according to site responses;
- Reach: the possibility of reaching most citizens without the limitations of technical features or special physical or cognitive needs;
- Physical effort: easiness to use the government websites, as a result of data reuse;
- Trust: demonstrating reliability and credibility to ensure security in the information exchange and site navigation.

Six additional components were added to the G-quality evaluation besides Nielsen's 10, namely accessibility, interoperability, security and privacy, information truth and precision, service agility, and transparency. Figure 6 illustrates the mapping between the evaluation criteria and the heuristic rules.

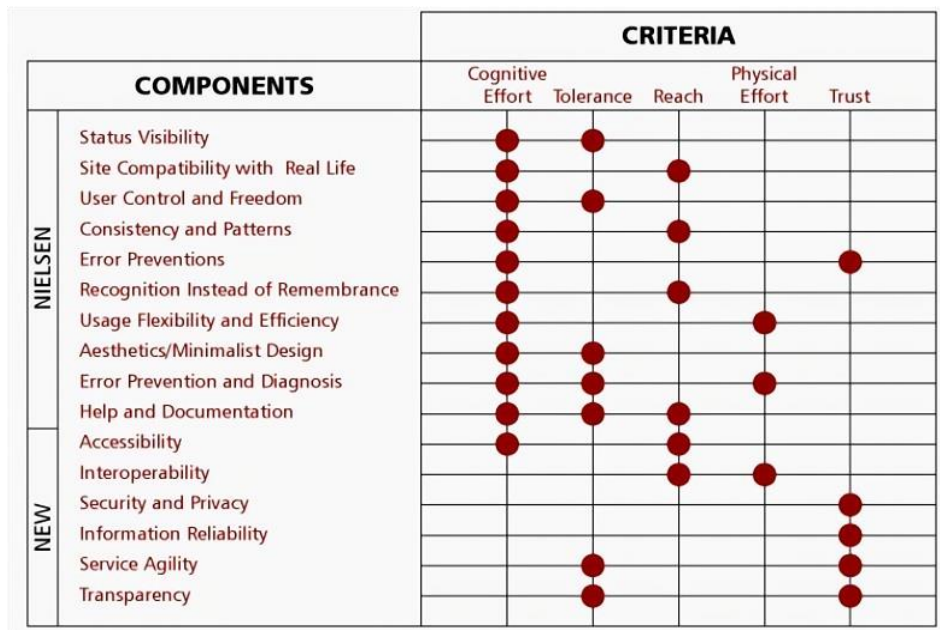

Fig. 6. The evaluation criteria and heuristic rules mapping (Garcia et al., 2005)

The G-Quality evaluation was also used to evaluate three similar e-government sites: an Ecuadorian website, a Latin-American website, and a European website (Granizo et al., 2011). Table 2 shows the heuristics set of the g-Quality evaluation method used in the study (Granizo et al., 2011).

Table 2: Heuristics of the g-Quality evaluation method (Granizo et al., 2011)

| ID | Heuristic |
|----|-----------|
| H1 | Visibility of system status |
| H2 | Match between the system and the real world |
| H3 | User control and freedom |
| H4 | Consistency and standards |
| H5 | Error prevention |
| H6 | Recognition rather than recall |
| H7 | Flexibility and efficiency of use |

| H8  | Aesthetics and minimalist design |
| --- | --- |
| H9  | Help users recognize, diagnose, and recover from errors |
| H10 | Help and documentation |
| H11 | Accessibility |
| H12 | Interoperability |
| H13 | Security and privacy |
| H14 | Information truth and precision |
| H15 | Service Agility |
| H16 | Transparency |

In Table 2, H1-H10 are based on Nielsen's ten heuristics, and H11-H16 are added. Granizo et al. (2011) summarize the main recurrent problems, including inadequate management of visited and unvisited links, outdated information, insufficient guides, lack of liberty to manage the process, and no place on the site for transparency and backup of the complaint. The study also showed that heuristic H4 (i.e., consistency and standards) was the most infringed heuristic related to the 16 critical problems identified, H1 (i.e., visibility of system status) was the second (Granizo et al., 2011).

Research-Based Web Design & Usability Guidelines (RBED&UG) were used in a recent evaluation study (Benaida, 2023). This study explored the usability of e-government websites between developing and developed countries (Algeria and the UK). In the study, three chapters (Page Layout, Text Appearance, Graphics, Images & Multimedia) from the RBED&UG were selectively examined as representatives of the other 15 chapters. Following the original guidelines' five-point scale, a consistent five-point scale was implemented in the questionnaire, ranging from "very poor" to "very strong." The questionnaires provided to experts focused on evaluating Algerian and British e-Government services concerning the three selected guidelines of usability. The results show that the Algerian e-government websites lack mostly in usability design, whereas the UK's performed well (Benaida, 2023). The study suggests that Algerian e-government representatives and developers should focus on these identified weaknesses to improve the usability of the systems. Emphasis should be placed on the weakest areas highlighted by the study, considering their relative importance. This study demonstrates the importance of usability design and evaluation during the development process of e-government.

**3.3   Accessibility**

Accessibility is defined as the extent to which products, systems, services, environments, and facilities can be used by a population with the widest range of user needs, characteristics, and capabilities to achieve identified goals in identified contexts of use (ISO 9241-112:2017, 2017). Table 3 summarizes the accessibility-related studies carried out across 96 countries, with the Americas ranking at the top (Cisneros et al., 2021).

Table 3: Accessibility research works by region (Cisneros et al., 2021)

| Region | Number of studies in that region |
|---|---|
| Americas | 42 |
| Africa | 37 |
| Asia | 27 |
| Europe | 24 |

The review also found that the Web Content Accessibility Guidelines 2.0 (WCAG 2.0) were primarily used in accessibility evaluations of e-government website applications (see Table 4) (WCAG 2.0, 2008).

Table 4: Accessibility guidelines used, by region (Cisneros et al., 2021)

| Guideline | Number of times used |
|---|---|
| Web Content Accessibility Guidelines 2.0 | 30 |
| Web Content Accessibility Guidelines 1.0 | 5 |
| Section 508 | 3 |
| Own Proposal | 2 |
| Accessibility Model of Electronic Government | 1 |
| Korean Web Content Accessibility Guidelines 2.0 | 1 |

WCAG defines a set of guidelines developed by the Web Accessibility Initiative (WAI) of the World Wide Web Consortium (W3C). WCAG inspects the accessibility of a website taking into account the layout, color choice, readability, and browser independence. These guidelines are defined to make web content more accessible to people with disabilities. The primary focus of WCAG is to ensure that web content is perceivable, operable, understandable, and robust for all users, regardless of their abilities or disabilities (WCAG 2.0, 2008). In July 2023, WCAG 3.0 Draft was published online. WCAG 3 provides recommendations that explain how to make the web more accessible to people with disabilities and apply the guidelines to web content, apps, tools, publishing, and emerging technologies on the web (WCAG 3.0 Draft, 2023).

Paul & Das (2020) found that very few websites adhere to WCAG standards for accessibility across 65 Indian e-government websites surveyed; they identified many accessibility errors and usability issues in the evaluation. During the e-government website design and development, less priority to accessibility and usability was given, resulting in inefficient service delivery, poor adoption, and user engagement. Thus, it is important to improve the overall accessibility of e-government websites.

### 3.4 Access to information

Access to information is a cornerstone of HCI in e-government and e-democracy. Ensuring that citizens have easy and equitable access to information is crucial for fostering transparency, promoting civic engagement, and enhancing the overall effectiveness of digital governance initiatives.

Access to information promotes transparency in government operations. Through well-designed UI and information architectures, citizens can access a wide range of public information, including policies, decisions, and government activities. Informed citizens are better equipped to participate meaningfully in democratic processes. HCI emphasizes user-centered design, ensuring that information is presented in a way that is easily understandable and navigable for diverse users. User-friendly UIs contribute to an inclusive environment, accommodating individuals with varying levels of digital literacy and diverse information needs.

For example, e-government improves the quality of public services and helps deliver information more effectively to the public. However, village government administrations have not supported e-government information systems and public services in developing countries such as Jambi Province in Indonesia (Maulana & Bafadhal, 2020). Currently, many examples show that village administrations are not fully integrated into robust database systems. Therefore, relevant government departments must support the e-government of villages to achieve information accessibility strongly (Maulana & Bafadhal, 2020).

Open Government Data (OGD) refers to the idea that specific government data should be made available to the public in a format that is easily accessible, downloadable, and usable. The concept revolves around the principle of transparency and aims to empower citizens, businesses, researchers, and other stakeholders by providing them with government-generated data (Çaldağ, 2019; Ferati et al., 2020). However, the user needs concerning open data remain unknown (Degbelo, 2020). Degbelo (2020) synthesized the needs of OGD users based on a critical review of empirically derived OGD usage issues in the literature and defined a preliminary classification method for user needs, as shown in Figure 7 (Degbelo, 2020). The review indicates that user needs concerning open data interaction are primarily informational and transactional, with less emphasis on navigational needs. Future work should delve deeper into these findings through additional user-based studies.

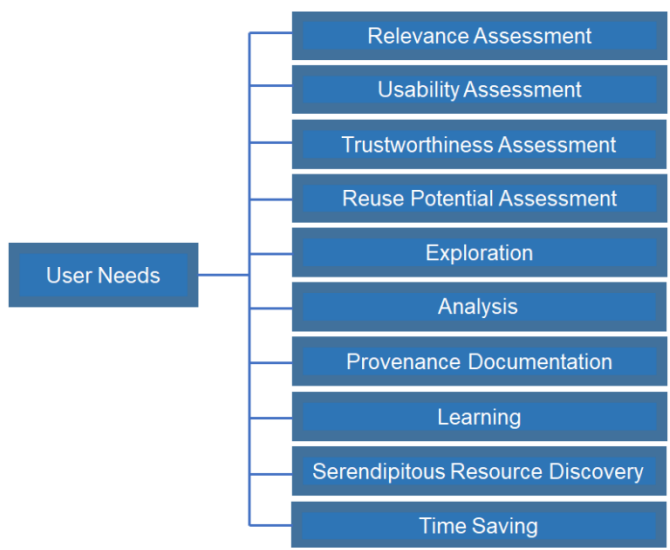

Fig. 7. Preliminary taxonomy of open data user needs (Degbelo, 2020)

## 3.5 Transaction efficiency

Transaction efficiency refers to the speed, accuracy, and ease with which citizens can complete transactions or tasks within e-government and e-democracy platforms. Citizens often engage in various transactions or interactions with e-government systems. These transactions can include activities such as accessing government services, submitting forms, participating in online consultations, or voting in electronic elections.

Transaction efficiency focuses on optimizing the design and functionality of the UI to streamline these processes, ensuring that users can accomplish their goals quickly and without unnecessary obstacles. Reducing the number of steps or clicks needed to complete a transaction can significantly improve efficiency. Users appreciate systems that allow them to achieve their objectives with minimal effort.

A website or system that conforms to HCI design principles can improve transaction efficiency and make user operations smoother and simpler. Conversely, unreasonable design can cause a poor user experience and affect the final completion result.

Eilu et al. (2021) studied the transaction efficiency of online tax services in Uganda and revealed that taxpayers underutilized online tax return services because of their poor usability design. Furthermore, the study found that most taxpayers found the electronic tax payment system more challenging than the others (Eilu et al., 2021). Therefore, Eilu et al. argued that the tax filing processes of online tax systems should be designed concisely with fewer operation steps, there is significant room for improvement in transaction efficiency, and HCI experts in Uganda could make substantial improvements (Eilu et al., 2021).

Karayumak et al. (2011) studied a Helios voting system, and they had previously proposed an improved version of the UI of the system, explicitly focusing on vote casting and individual verifiability, aiming to improve the transaction efficiency of the remote e-voting system. The study tested the UI in a simulated mayoral election and revealed that the UI was improved to support transaction efficiency (Karayumak et al., 2011).

### 3.6 Security and privacy

A secure and private digital environment fosters trust in government and democratic processes. Citizens are more likely to engage with e-government services and participate in e-democracy initiatives when they trust that their information is handled carefully. E-government systems and platforms are attractive targets for cyber threats. Ensuring robust security measures helps prevent cyber attacks, data breaches, and unauthorized access, safeguarding critical infrastructure and sensitive information.

Conventional usability testing methods may not be sufficient when security and privacy are involved, especially considering the likelihood of non-optimal security configurations by casual users prioritizing other matters over security. Therefore, there is a need for a new approach to usability testing that specifically addresses usable security issues, aiming to outline and advocate for this method in assessing the usability of security tools and techniques (Greitzer, 2011).

Khairnar & Kharat (2016) surveyed secure online voting systems to examine various online voting systems based on techniques such as homomorphic encryption and blind signatures. Building upon this analysis, the researchers presented a novel and user-friendly secure online voting system comprising biometric authentication. Unlike previous systems where casted votes were not stored securely or separately, the proposed system ensures robust security by employing advanced encryption methods and restricting voting to authenticated individuals. The system was designed to provide a high level of security for online voters, safeguarding them against diverse security threats (see Figure 8). With reliability in both casting and recording votes, this system emerged as a secure and trustworthy solution, particularly suitable for remote voters (Khairnar & Kharat, 2016).

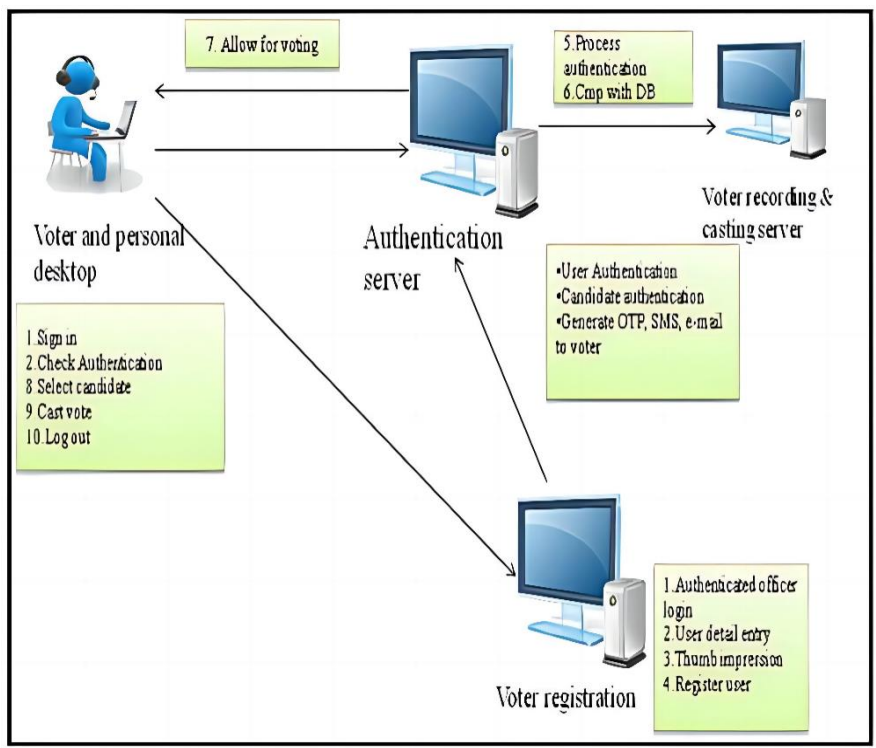

Fig. 8. Architecture of Online Voting System(Khairnar & Kharat, 2016)

While the concept of "privacy" has been thoroughly explored in regulatory and scholarly discussions, Wong & Mulligan (2019) highlighted a relative lack of attention to the intricate aspects of "design" within HCI. Through a literature review of HCI publications on privacy and design, they identified dimensions that connect design to privacy, including the purpose of design, the actors involved, and the intended beneficiaries. The researchers proposed new roles for HCI and design in PBD (privacy by design) research and practice, advocating for values- and critically-oriented design approaches. These approaches emphasize social values, shaping the definition of privacy problem spaces (Wong & Mulligan, 2019).

As data security and privacy concerns grow, HCI will play a crucial role in designing UI that prioritize user trust. Clear communication about data usage, robust authentication methods, and privacy-preserving design will be crucial to the success of e-governance and e-democracy.

In that respect, driven by the HCI design principles and methods, HCI work offers design implications that can contribute to optimizing e-government and e-democracy. HCI work can identify problems in existing websites and systems, make targeted improvements based on specific issues, pay more attention to the needs of citizens, and improve the efficiency of e-government and e-democracy through design changes. Even if many websites and systems do not consider HCI in the initial design stage, HCI will still play a significant role in the subsequent iteration process for improvement.

## 4 Case studies

To illustrate the design implications and contributions of HCI work to e-Government and e-Democracy, this section presents three case studies.

### 4.1 HCI in e-government

E-government in China has achieved high-quality and rapid development, and Shanghai's e-government website is an example (refer to sh.gov.cn). Shanghai's "Yi Wang Tong Ban" ("一网通办" in Chinese and "Government Online-Offline Shanghai" in English) is a comprehensive government service platform offering convenient online services to residents. Through this platform, citizens can effortlessly handle various governmental affairs without physically visiting government offices. The services provided encompass a wide range, including but not limited to residency permit applications, traffic violation processing, social insurance inquiries, and medical service appointments.

Key features of "Yi Wang Tong Ban" include:

- Comprehensive Services: Integrating services from multiple government departments, "Yi Wang Tong Ban" allows citizens to accomplish various tasks on a single platform, streamlining the often intricate processes;
- Online Processing: Citizens can submit applications online anytime, eliminating the need for physical presence, queues, and time wastage;
- Transparent Information: The platform provides detailed guides and process explanations, enhancing the transparency of each service's procedures and allowing citizens to understand the steps involved;
- Digital Services: "Yi Wang Tong Ban" offers features such as electronic certificates and online payments by leveraging digital technologies, simplifying various governmental processes for citizens.

"Yi Wang Tong Ban" aims to elevate the efficiency of government services, providing residents with a more convenient and expeditious experience for their administrative needs. It can be clearly seen that there is a function that supports language switching between simplified Chinese, traditional Chinese, and English in the upper right corner of the website homepage (see Figures 9 and 10).

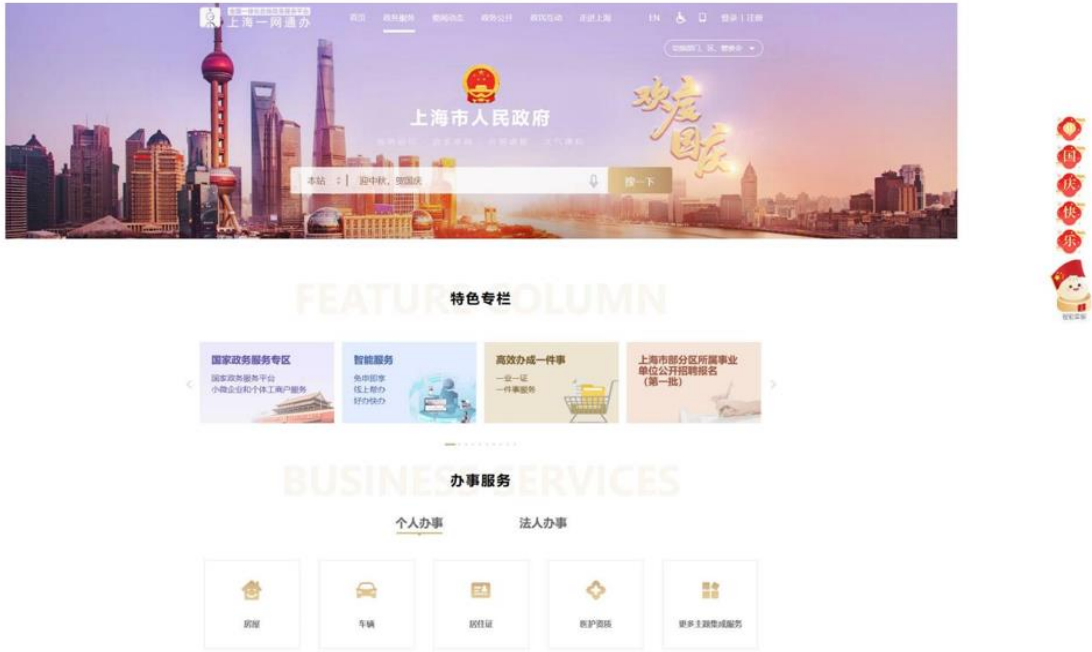

Fig. 9. The "Yi Wang Tong Ban" homepage (the Chinese version)

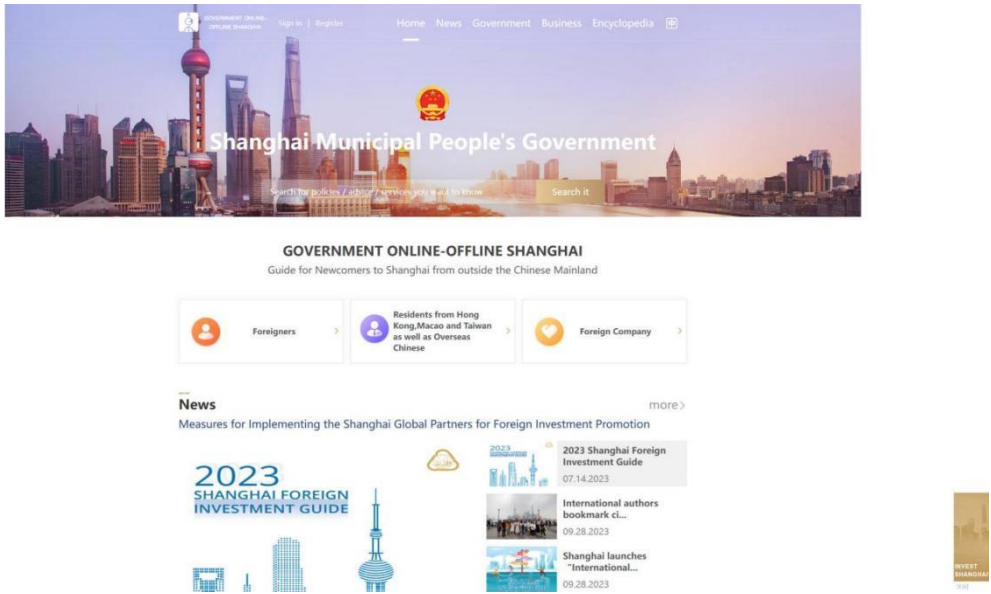

Fig. 10. The "Yi Wang Tong Ban" homepage (the English version)

An assessment of the website shows the following findings according to the six HCI design principles discussed in Section 3.

- Citizen-centered design: For Chinese users, the homepage primarily presents more options for transactions, whereas for English users, the homepage predominantly showcases introductions to the city and news. This design is based on an analysis of user needs. For example, displaying professional qualifications, document processing, transportation information, and healthcare in personal affairs aligns with the majority of citizens' needs. Placing them in prominent positions makes the operation more convenient;

- Usability: The UI of "Yi Wang Tong Ban" adheres to Nielsen's ten Usability Heuristics for User Interface Design and demonstrates good performance in terms of the five evaluation criteria (i.e., cognitive effort, tolerance, reach, physical effort, and trust). The usability design can be highlighted as follows: consistent design, simple and unified design style, appealing visual aesthetics design, no redundant information, a simple and clear information hierarchy, personalization design, and a low cognitive load on users;

- Accessibility: There are also accessible and caring versions in the header, demonstrating that the interface design fully considers persons with disability and elderly people (see Figure 11). By switching to the accessible version, citizens can choose targeted options such as sound, color schemes, and cursor shapes, which are comprehensive and meet the needs of different citizen groups. The caring version for elderly people shares similar features as the accessible version with the option of choosing larger font sizes for elderly people;

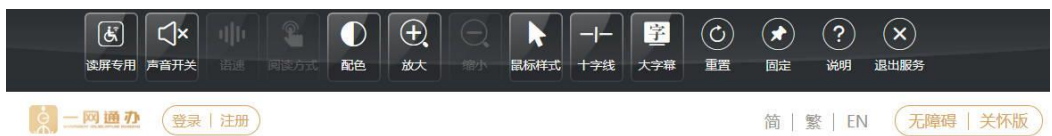

Fig. 11. The accessible functions of the "Yi Wang Tong Ban" website

- Access to information: It is very easy for citizens to access public information on the homepage with a comprehensive and intuitive information architecture (see Figure 12);

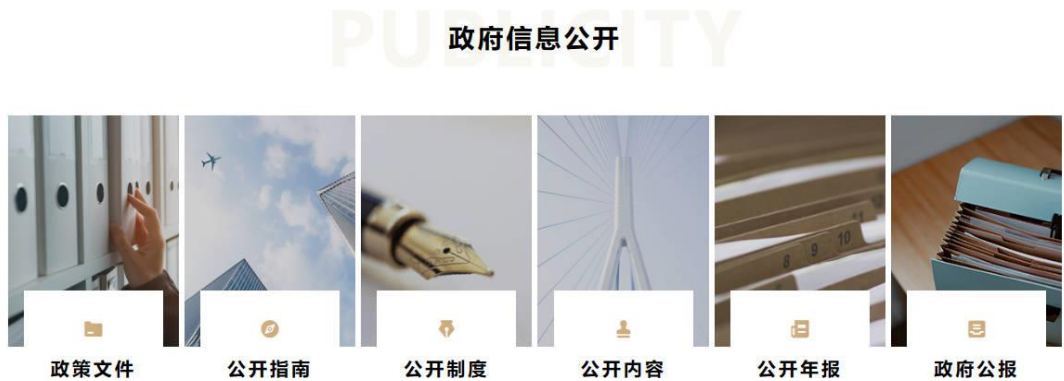

Fig. 12. Yi Wang Tong Ban Access to information

- Transaction efficiency: Transaction efficiency was considered in the design. For example, to operate issuing a certificate, users can directly search or click on the corresponding icon and download the electronic version without too many complicated operation steps;
- Security and privacy: The strict registration process and confirmation of login information ensure citizens' privacy and the website's security. The relevant government departments attach great importance to the privacy and security of websites by following the national regulations on security and privacy (Reddick & Zheng, 2018).

Chen et al. (2022) initially assessed the "Yi Wang Tong Ban" website. They found that there are issues with insufficient user engagement, limited participation channels, and low levels of participation. Although there is a high proportion of users who responded as "very satisfied" or "satisfied" in the survey, most users took the survey at the request of staff, leading to the phenomenon of "high praise and difficult to find bad reviews," making it challenging to collect objective feedback from the users.

In summary, based on the analysis of the six HCI principles in e-government, the "Yi Wang Tong Ban" website demonstrates a reasonably good design example. However, a third-party or anonymous evaluation should be conducted to collect more objective user feedback. The improvement of an e-governance website cannot be achieved solely through design principles and guidelines; feedback from users/citizens is critical.

## 4.2　HCI in e-democracy

E-voting is an excellent example of e-democracy. Acemyan et al. (2022) implemented a STAR-Vote project to create a highly usable and secure end-to-end (e2e) voting system. STAR-Vote hides complex security mechanisms (see Figure 13), thus conforming to the mental model of voters' paper voting. The specific voting process is shown in Figure 14. Simple and plain language, clear instructions, and intuitive interfaces matched typical how-to-vote procedures (see Figure 15). This design preserves the security of the voting system while also considering its usability.

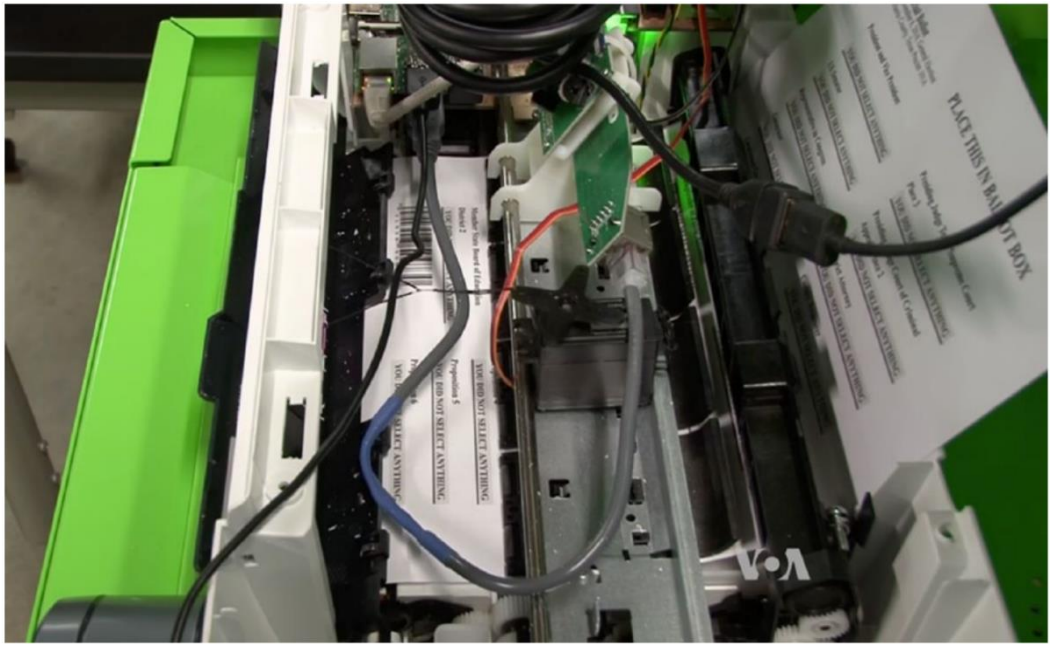

Fig. 13. The ballot box's scanner and paper-diverting mechanism (Acemyan et al., 2022)

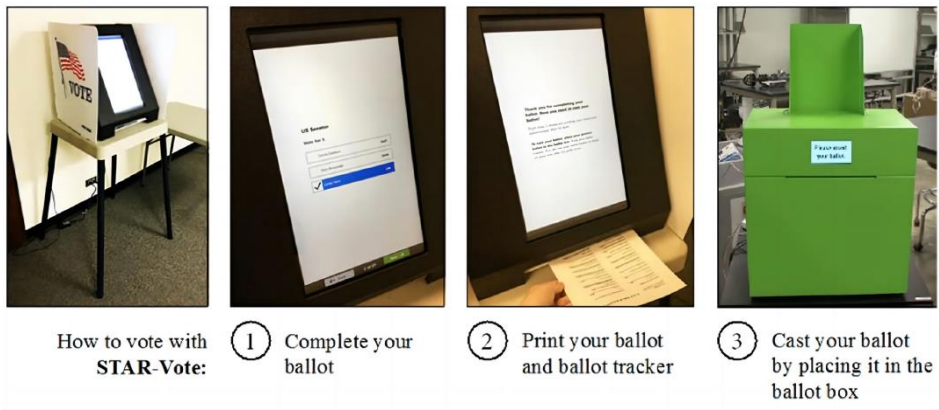

Fig. 14. STAR-Vote's typical how-to-vote procedures (Acemyan et al., 2022)

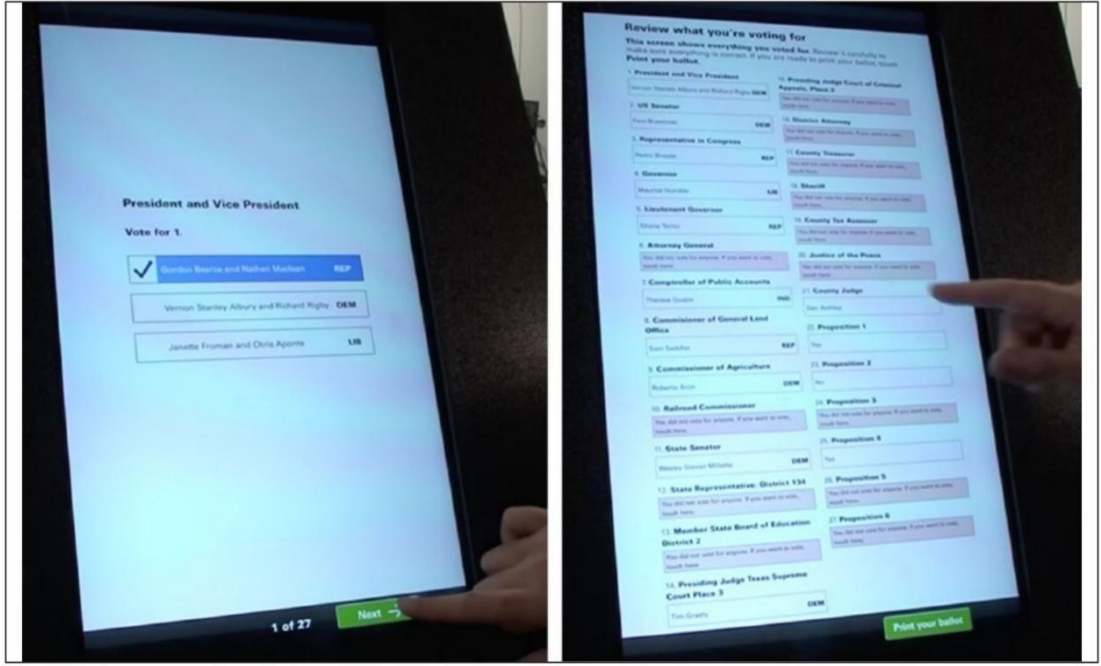

Fig. 15. STAR-Vote's ballot completion interfaces (Acemyan et al., 2022)

From a design process perspective, a user-centered design approach was employed through collaboration among computer security experts, auditors, human factors researchers, and election officials. The user-centered design process for STAR-Vote began with an interdisciplinary team comprising security researchers, statistical auditors, human factors experts, and election officials. The team held meetings to establish a consensus on the architecture and specifications of STAR-Vote, emphasizing the equal prioritization of usability and security. In this initial phase, human factors experts, drawing from over a decade of experience in studying voting, utilized established HCI heuristics and findings from prior human factors research. They compiled a comprehensive list of UX properties to be integrated into the system. Examples of these properties included presenting one race at a time, using simple language understandable to all voters, incorporating large and visible navigation buttons, allowing voters to touch their selections, notifying voters of incomplete selections, and enabling direct navigation from the review screen to specific races on the ballot with a touch. The project underwent iterative design, including human factors research and usability testing, to ensure both usability and security.

A summative system usability assessment was conducted following ISO 9241-11 criteria. The results indicated that STAR-Vote is the most usable and cryptographically secure e2e voting system compared to previously tested systems. The conclusion highlights the significance of STAR-Vote as a tamper-resistant voting system for the U.S. elections, ensuring both electoral integrity and accurate representation of voter intent. Specifically, the study shows that celebrity voting outperforms the other three (Helios, PaV, and Scantegrity) regarding voting time and percentage of successful votes. An important reason for the success of celebrity voting is to consider usability issues in the early stages of design rather than waiting for the system to be designed before discovering existing issues through usability testing.

Previous studies indicate that complex security mechanisms could significantly impact the system's usability (Tognazzini et al., 2005). However, this project implies that a complex and secure system can still be highly usable, debunking the notion that implemented security compromises usability (Acemyan et al., 2022).

### 4.3    E-Estonia

Estonia has pioneered the development of e-government, earning a reputation as one of the most digitally advanced nations (Anthes, 2015; Goede, 2019; Tupay, 2020). EGDI and OSI (Online Services Index) of Estonia rank high in the VH (very high) rating class (15 countries) of the EGDI group(see Table 5).

Table 5: Leading countries in e-government development. (United Nations E-Government Survey 2022)

| Country name | Rating class | Region | OSI | EGDI(2022) |
|---|---|---|---|---|
| Estonia | VH | Europe | 1.0000 | 0.9393 |
| Finland | VH | Europe | 0.9833 | 0.9533 |
| Republic of Korea | VH | Asia | 0.9826 | 0.9529 |
| Denmark | VH | Europe | 0.9797 | 0.9717 |
| Singapore | VH | Asia | 0.9620 | 0.9133 |
| New Zealand | VH | Oceania | 0.9579 | 0.9432 |
| Australia | VH | Oceania | 0.9380 | 0.9405 |
| United States of America | VH | Americas | 0.9304 | 0.9151 |
| Japan | VH | Asia | 0.9094 | 0.9002 |
| Netherlands | VH | Europe | 0.9026 | 0.9384 |
| United Arab Emirates | VH | Asia | 0.9014 | 0.9010 |
| Sweden | VH | Europe | 0.9002 | 0.9410 |
| Iceland | VH | Europe | 0.8867 | 0.9410 |
| United Kingdom of Great Britain and Northern Ireland | VH | Europe | 0.8859 | 0.9138 |
| Malta | VH | Europe | 0.8849 | 0.8943 |

"E-Estonia"(https://e-estonia.com/) is a comprehensive initiative by the Estonian government to transform the country into a digitally advanced society. It encompasses several aspects of digital governance and services, such as e-identity, cyber security, and interoperability services (see Figures 16 and 17), making Estonia a global leader in e-government.

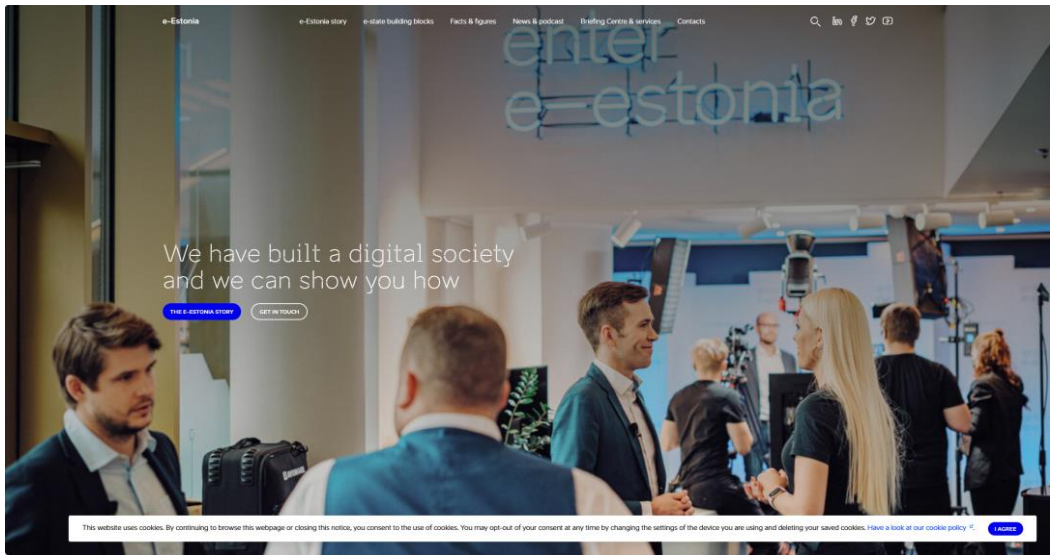

Fig. 16. The home page of the e-Estonia website (https://e-estonia.com/)

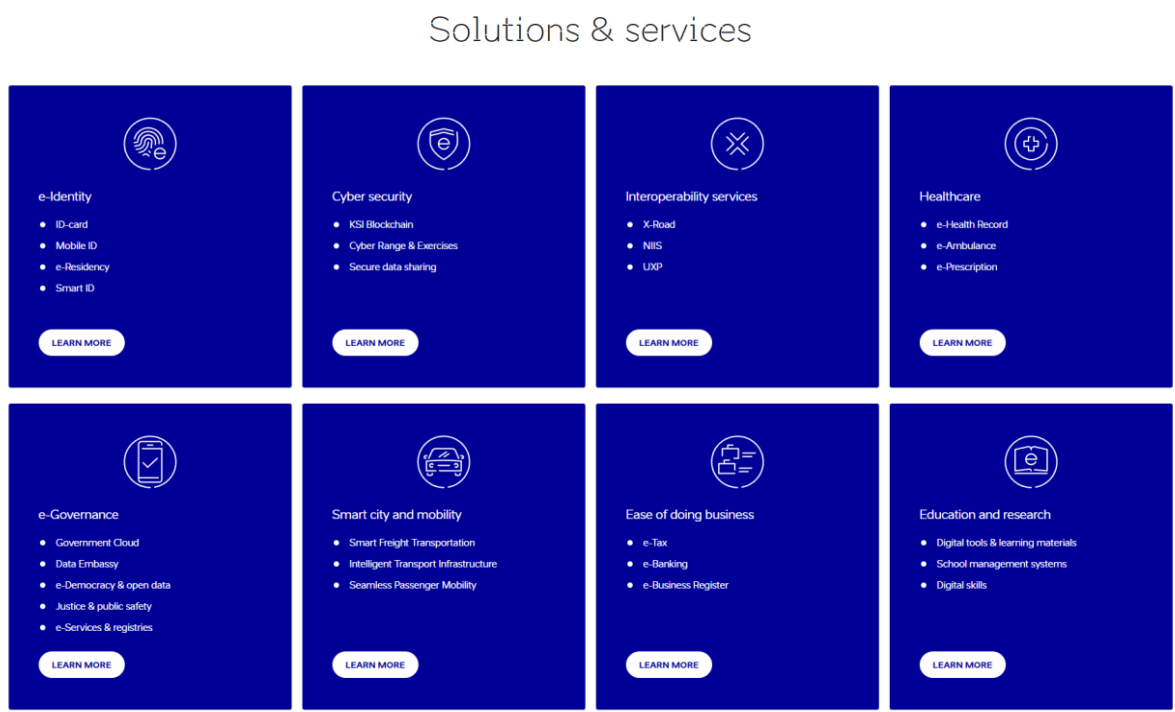

Fig. 17. Solutions and services offered on the e-Estonia website ( https://e-estonia.com/solutions/ )

Estonia's advanced e-governance infrastructure allows various government agencies and other organizations to share data seamlessly while ensuring privacy and security. For example, in 2001, the X-Road infrastructure facilitated secure data exchange between government databases and e-service platforms (see Figure 18). It enables secure and standardized data exchange between different information systems. X-Road is associated with As shown in Figure 18, "X-Road" has become the backbone of e-Estonia, allowing the nation's public and private sector information systems to link up and operate in harmony, resulting in 99% of public services being accessible online 24/7".

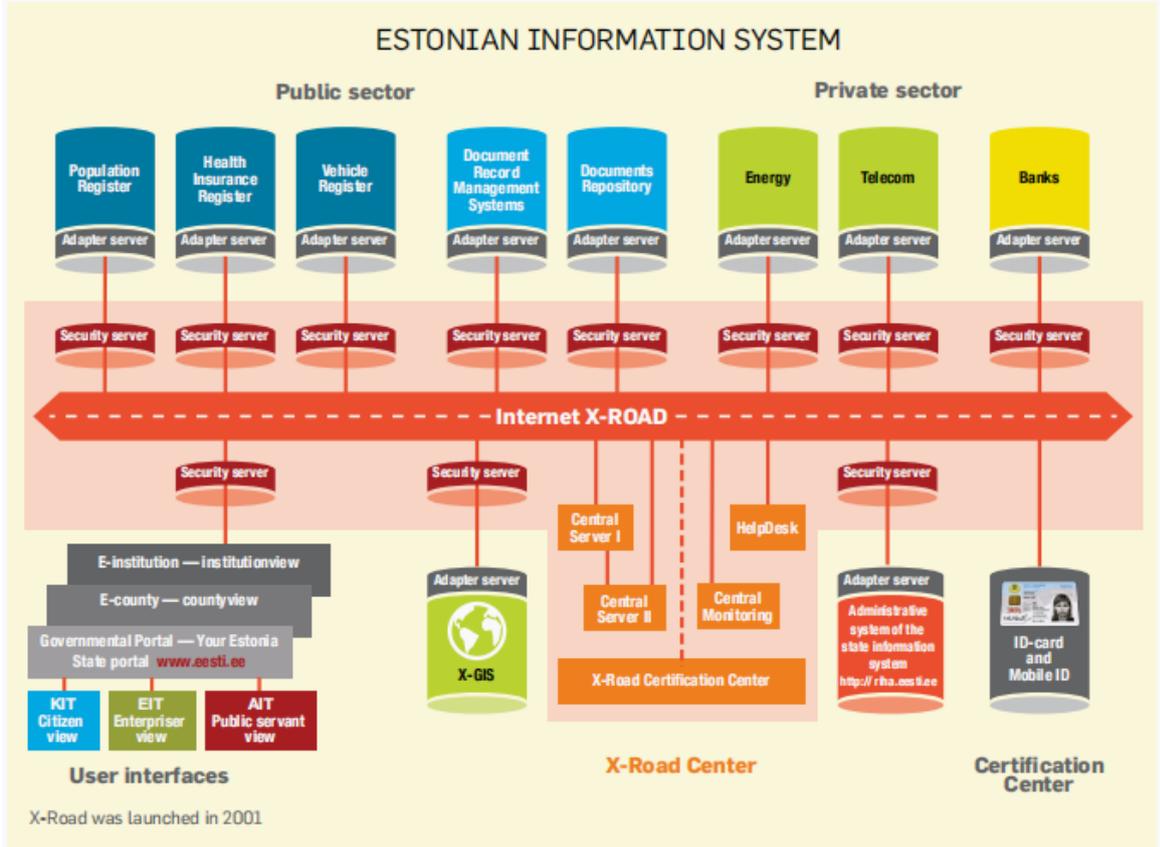

Fig. 18. Illustration of the databases and other architectural components of the x-Road e-Estonia (Anthes, 2015).

As a milestone event, Estonia's e-Residency Program is a groundbreaking initiative that allows individuals from anywhere in the world to become virtual residents of Estonia, gaining access to the country's digital infrastructure and e-services (Kotka et al., 2015; Tammpuu & Masso, 2019). However, convenient registration brings potential criminal risks. To address the potential risks, Sullivan & Burger (2017) suggested blockchain technology could provide adequate security guarantees and revolutionize the control and authentication of identity information.

Similar to the previous two case studies, we analyze e-Estonia by following the six HCI design principles:

- Citizen-centered design: E-Estonia considers all aspects of citizens' social lives. It was designed with the user in mind, making government services more accessible almost all matters can be resolved online. For example, e-Marriage services allow citizens to submit marriage applications online via the e-population register. In 2023, e-Estonia State App was built with the user in mind, making government services more accessible and convenient;
- Usability: The UI of "E-Estonia" adheres to the usability requirements defined in the g-Quality evaluation method. The usability design can be characterized as follows: exquisite and consistent design, concise and clear design style, simple information hierarchy and navigation design, and low cognitive load on users;
- Accessibility: There is no alternative version available like the One Network Service, but the UI for information download provides documents in different languages, including English, French, German, Spanish, and Japanese;
- Access to information: Citizens can easily access and download information on the E-Estonia website, which provides comprehensive and detailed information;
- Transaction efficiency: The transaction is very efficient, without complex operational steps. For foreign users, they can quickly accomplish tasks using the e-resident service and access registration instructions when needed;
- Security and privacy: One major advantage of the e-government platform is the application of blockchain technology to ensure the security of citizens' data and privacy. As mentioned earlier, the data within E-Estonia is distributed and stored in an encrypted form.

## 5    General challenges

Although HCI has played a crucial role in the successful implementation of e-government and e-democracy, there are still challenges to be addressed. This session highlights certain general challenges.

### 5.1    Ethical considerations

Ethical issues in e-government and e-democracy mainly include trust, privacy and security, transparency, and social media (Gajendra et al., 2012; Dwivedi et al., 2017).

Trust is a delicate aspect of the user-government relationship, and the design and presentation of e-government websites play a pivotal role in building and maintaining that trust. A cross-cultural e-government quality study (Aladwani, 2013) revealed that using language familiar to users and following real-world conventions in presenting information on e-government websites is crucial. This ensures that users can easily understand the content and navigate the site. Huang & Benyoucef (2014) found a close correlation between usability and credibility, as e-government websites with high usability were perceived as having higher credibility, and vice versa (Huang & Benyoucef, 2014).

Security, privacy, and interoperability are the key elements to building trust among users in e-government systems (Manda & Backhouse, 2016). Protecting citizen privacy is also essential in building trust in e-government initiatives and development. Privacy can be promoted by setting up guidelines, procedures, compliance programs, and training and awareness. Privacy can also be enhanced by applying technologies such as blockchain, artificial intelligence, and machine learning (Yang et al., 2019).

Dwivedi et al. (2017) suggested that governments should widely adopt open data practices to promote transparency. Governments are encouraged to adopt open data practices by exposing all their processes, including bidding, contracting, form processing, public monitoring of activities, and employee activities. They also emphasize the significance of social media in e-government, particularly through e-participation. Social media is identified as a crucial tool for governments, enabling e-participation. It facilitates the redistribution of power by involving citizens traditionally excluded from government decisions and policy-making (Dwivedi et al., 2017). While social media has the potential to be a powerful tool, there is evidence suggesting that it is not always successful in achieving its intended outcomes (Harrison et al., 2011). Before implementing recommendations from public participation, fundamental questions such as who participates, how participants exchange information and make decisions, and the basis for incorporating suggestions into decision-making processes need to be addressed (Dwivedi et al., 2017).

## 5.2 Multilingualism and multicultural issues

Multilingualism in e-government is not just a matter of linguistic convenience; it is a strategic imperative that promotes equity, inclusivity, and effective communication, ultimately contributing to the success and legitimacy of e-government initiatives. Kocór et al. (2017) explored enhancing the quality of information systems to support democracy and public administration in Slavic countries between Western Europe and Russia. The proposed approach involves using a constructed language designed explicitly for this region. This language is intended to serve as a viable alternative to English, aiming to elevate the overall quality of Information and Communication Technology (ICT) used in e-democracy initiatives. The underlying assumption is that language, education, and e-democracy form a developing triad (Kocór et al., 2017).

Aldrees & Gračanin's study (2021) emphasizes the necessity of addressing multicultural requirements in the design of e-government services. Specifically, the study focused on the multicultural usability of the Saudi job-seeking web portal ("Taqat") in comparison to the United States (US) job-seeking web portal ( "USAJOBS") from the perspectives of Saudi and US citizens. The assessment evaluated both portals' compatibility with US web design guidelines (USWDS) (U.S. Government, 2022). The Saudi web portal is strongly aligned with US design guidelines, lacking only three out of nineteen. A follow-up user study was conducted with 200 participants, 100 Saudi and 100 US citizens, using a web-based survey with twenty close-ended questions measured on a five-point Likert scale and Nielsen's ten heuristic usability principles. The results, analyzed using the Mann-Whitney U test, indicated a significant multicultural gap in the usability of the Saudi web portal compared to the US counterpart. Moreover, future efforts and investigations must mitigate these multicultural differences in the usability and adoption of the Saudi job-seeking web portal (Aldrees & Gračanin, 2021).

## 5.3  Digital divide

Digital divide refers to the unequal access of citizens to information and communication technologies (ICT) and the uneven possession of skills and experience required for using it. Analyzing panel data from 27 European countries spanning 2010 to 2018, Pérez-Morote et al. (2020) identified key factors influencing citizens' use of e-government services. These factors include evaluations of e-government services from the supply side, citizens' trust in governments, and the digital divide related to income and education levels.

The digital divide in e-government extends the concept to disparities in accessing and utilizing government services through digital platforms. It encompasses differences in citizens' ability to access, use, and benefit from online government services due to variations in technology access, digital literacy, and other socio-economic factors.

## 5.4 The balance between innovations and risks

Improving the transparency of e-government websites is beneficial for citizens to access information and enhance the usability design of these websites, but excessive transparency carries potential risks. The intersection of transparency, e-government, and the potential risks associated with the careless handling of personal information leads to a phenomenon called "hyper-transparency". While transparency and e-government aim to foster interaction between citizens and public institutions, mishandling personal information can result in an unreasonable and excessive disclosure of data, causing distortions in society (Rodríguez-Hoyos et al., 2018).

Big data analytics support democratic processes by enhancing efficiency, effectiveness, and transparency for governments and organizations. However, automatic decision-making may lead to discrimination and compromising equality, a fundamental democratic value. The challenges of privacy threats for e-democracy may facilitate manipulation, polarisation, and disinformation(Mavriki & Karyda, 2022).

## 5.5 Data protection for e-democracy

Codreanu (2021) argued that integrating e-democracy and cybersecurity is crucial for future e-democracy from the data protection perspective. Neglecting cybersecurity can expose vulnerabilities that malicious cyber operations may exploit, jeopardizing the integrity of e-democracy. Addressing cybersecurity concerns protects against potential threats and fosters trust in both e-democracy and traditional democratic processes (Codreanu, 2021). As discussed in Section 2.2, e-democracy is not a substitute for conventional forms of democracy but a means to make democratic institutions more efficient and productive. Protecting e-democracy is about safeguarding the implementation of democratic systems.

As an example, e-voting is an essential component of e-democracy, determining the political direction of a country or region. Malicious network attacks are always unavoidable, so the data security of electronic democracy is critical. The case study discussed in Section 4.2 introduced the encryption method of the STAR-Vote, which can be considered a widely applicable method in the future. Of course, the mature use of blockchain technology, as used in e-Estonia, is also an effective way to protect the security of electronic democratic data.

## 6 Future trends

### 6.1 Personalization

HCI will increasingly prioritize citizen-centered design, creating intuitive, accessible, and inclusive UIs that meet diverse user needs. The emphasis will be on enhancing the overall user experience. There will be a growing trend toward personalized UIs in e-government and e-democracy platforms. Tailoring experiences to individual preferences can improve engagement and satisfaction. Personalization deepens citizen-centered design, providing more targeted services to citizens while grasping their needs.

### 6.2 Enhanced security with technologies

Blockchain technology has the potential to significantly impact HCI in e-government and e-democracy by introducing trust, transparency, and security into digital transactions and processes. Blockchain provides a decentralized and secure ledger for transactions. In e-government, this could mean more transparent and tamper-resistant record-keeping, enhancing trust in digital interactions.

For example, blockchain can be applied to e-democracy for secure and transparent voting systems. Each vote is recorded on the blockchain, ensuring that it cannot be altered or manipulated, addressing concerns related to the integrity of the electoral process (Pawade et al., 2020). Blockchain can also offer a decentralized and secure solution for identity management. Citizens can have control over their personal data, and this decentralized identity could be used in various e-government services, enhancing privacy. Also, biometric technology can secure online voting systems (Pawade et al., 2020).

### 6.3 Globalization

HCI plays a role in cultivating a sense of global digital citizenship. E-democracy platforms can encourage citizens to think beyond national borders, fostering a sense of shared responsibility for global issues and encouraging international collaboration. Considering the global audience, multilingual interfaces become essential. E-government platforms should provide content and services in multiple languages to cater to citizens with different linguistic backgrounds. Designing user and other interfaces that transcend geographical and cultural boundaries while addressing global challenges is essential for creating a connected and participatory digital world.

## 6.4 Towards AI-government and AI-democracy

Society is transitioning from a computer to an AI era (Xu, Dainoff, et al., 2023). We argue that e-government and e-democracy are moving towards AI-government and AI-democracy. The future trends in e-government and e-democracy are increasingly intertwined with the integration of AI. AI will play a crucial role in automating routine government services. Chatbots and virtual assistants powered by AI will handle inquiries, provide information, and assist citizens in completing various transactions. Vrabie (2023) introduces the concept of e-government 3.0, which builds on e-government 2.0 principles and emphasizes the use of emerging technologies, particularly AI, to revolutionize public service delivery and governance. In the future, integrating GPT-4 into the petition system and enhancing it with advanced machine-learning (ML) techniques can indeed lead to significant improvements (Vrabie, 2023).

AI-driven predictive analytics will enable governments to anticipate citizen needs and preferences. This can lead to more proactive and personalized service delivery, enhancing overall efficiency. AI can help analyze citizen feedback and sentiment. Natural Language Processing (NLP) can help governments understand public opinions, allowing for more responsive and citizen-centered decision-making. AI will contribute to advanced security measures in e-government, detecting and preventing cybersecurity threats. Machine learning algorithms can continuously adapt to evolving risks, ensuring robust data protection. Governments will increasingly rely on AI to analyze large datasets for data-driven decision-making. This can lead to more effective policy formulation and resource allocation.

The application of Natural Language Processing (NLP) in Bulgaria is a good example (Hristova et al., 2022). The study proposed an integrated ML-based AI system to support decision-makers in e-government and public service provision. The system draws on diverse data sources, including news websites, web forums, and other social networks. By adopting these advanced computational methods, the research aims to provide a more dynamic and intelligent means of understanding public sentiments towards e-government services, thereby contributing to informed decision-making and continuous improvement in service delivery (Hristova et al., 2022). Also, Al-Mushayt (2019) proposed a framework that utilizes AI technologies to automate and facilitate e-government services (Al-Mushayt, 2019). Leveraging trustworthy AI techniques to advance e-government services has the potential to bring about significant improvements in efficiency, cost-effectiveness, and citizen satisfaction.

While integrating AI in e-government and e-democracy holds immense potential, it also raises challenges such as ethical AI, privacy, and accountability. Striking a balance between technological innovation and responsible governance will be vital in realizing the future benefits of AI-government and AI-democracy.

## 7 Discussions and Conclusions

As discussed above and demonstrated by the three case studies, HCI has played a pivotal role in successfully implementing e-government and e-democracy initiatives. Successful e-government and e-democracy platforms take users and user needs as priorities driven by the user-centered design HCI approach. HCI methodologies, such as user interviews, persona-driven design, user interface prototyping, and usability testing, ensure that systems are intuitive and tailored to the needs and preferences of the users. HCI and its design principles have been instrumental in creating e-government and e-democracy systems that meet user needs with optimal UX. The advantages HCI provides allow for effective problem-solving in applying e-commerce paradigms, contributing to the success of e-government and e-democracy.

Although HCI has played a crucial role in the successful implementation of e-government and e-democracy, there are still challenges to be addressed in future work, such as ethical considerations, multilingualism and multicultural issues, digital divide, the balance between innovations and risks, and data protection for e-democracy. Looking forward, HCI will continuously contribute to e-government and e-democracy in many aspects, including personalization and enhanced security with advanced technologies such as blockchain, globalization, and AI-government and AI-democracy.

To address these challenges and support future work of e-government and e-democracy, the UCD approach promoted by HCI remains essential to ensuring that the design and development of e-government and e-democracy systems are based on user needs, preferences, and capabilities. In many cases, we need to challenge the existing approaches being taken. For example, Deibert (2018) proposed a "human-centric" alternative to challenge the prevalent "national security-centric" approach to cybersecurity. Instead of prioritizing state interests, this approach emphasizes human rights for access to information. As technologies advance, HCI research and application will augment new technologies to maximize their benefits to e-government and e-democracy.

There are research opportunities for HCI to make further contributions to e-government and e-democracy. In the context of emerging digitization and globalization, many research questions must be addressed. For example, have citizen needs for e-government and e-democracy undergone significant changes in some aspects compared to the past? Should we preserve citizen's traditional habits while adopting e-government and e-democracy systems? How can citizens accept new technologies while we apply them? From an HCI methodology perspective, HCI methods need to evolve. We need to develop innovative HCI methods and update HCI design principles to support future work. Just as the g-Quality evaluation method extended Nielsen's ten heuristics based on existing HCI design principles, new HCI design principles and research methods may emerge.

For HCI practitioners, new technologies (e.g., AI) provide new opportunities for e-government and e-democracy, challenge the current HCI approaches, and push the advancement of HCI approaches. HCI practitioners must keep up with the development of new AI technologies and understand the latest AI technologies to discover how technologies will impact e-government and e-democracy. From a process perspective, HCI practitioners need to actively participate in developing e-government and e-democracy systems to increase their influence by leveraging their unique expertise. HCI practitioners also need to promote interdisciplinary collaboration by sharing HCI knowledge with other professionals and integrating HCI approaches into the development process.

In summary, HCI plays a crucial role in developing e-government and e-democracy systems. HCI remains an essential approach to addressing future challenges, and there are also many opportunities for HCI, especially when new technologies are applied to e-governance and e-democracy systems. It also requires HCI practitioners to actively develop new methods and participate in the development process of these systems through interdisciplinary collaboration. Future e-government and e-democracy require HCI's further participation and contributions.